\documentclass[%
 reprint,
 amsmath,amssymb,
 aps,
prx,
]{revtex4-2}

\usepackage{braket}
\usepackage{mathbbol}
\usepackage{graphicx}
\usepackage{dcolumn}
\usepackage{bm}

\begin{document}

\preprint{APS/123-QED}

\title{Coherent Photogalvanic Effect for Second-Order Nonlinear Photonics}

\author{Ozan Yakar$^1$}
\author{Edgars Nitiss$^1$}%
\author{Jianqi Hu$^1$}%
\author{Camille-Sophie Br\`{e}s$^{1,}$}%
 \email{camille.bres@epfl.ch}
\affiliation{$^1$\'Ecole Polytechnique Fédérale de Lausanne (EPFL), Photonic Systems Laboratory (PHOSL), Lausanne CH-1015, Switzerland }

\date{\today}

\begin{abstract}
The coherent photogalvanic effect leads to the generation of a current under the absorption interference of coherent beams and allows for the inscription of space-charge gratings leading to an effective second-order susceptibility ($\chi^{(2)}$). The inscribed grating automatically results in quasi-phase-matching between the interfering beams. Theoretical and experimental studies have been carried out, mostly focusing on the degenerate case of second-harmonic generation, showing significant conversion efficiency enhancements. However, the link between the theory and experiment was not fully established such that general guidelines and achievable conversion efficiency for a given material platform are still unclear. In this work, we theoretically analyze the phenomenological model of coherent photogalvanic effect in optical waveguides. Our model predicts the existence of non-degenerate sum-frequency generation quasi-phase-matching gratings, which is confirmed experimentally for the first time. In addition, we rigorously formulate the time dynamics of the space-charge grating inscription in coherent photogalvanic process. Based on developed theoretical equations for the time dynamics of the space-charge grating formation, we extract the material parameters governing the process for our experimental platform, stoichiometric silicon nitride. The results obtained provides a basis to compare the performances and potentials of different platforms. This work not only supplements the theory of coherent photogalvanic effect, but also enables us to identify critical parameters and limiting factors for the inscription of $\chi^{(2)}$ gratings.
\end{abstract}

\maketitle


\section{INTRODUCTION}
All-optical control of currents in centrosymmetric media has been a widely pursued objective for over half a century \cite{manykin1967one}. Caused by the quantum interference of multiphoton absorption, coherent currents facilitate several physical and chemical processes \cite{manykin1967one, Jackson1983resonance, dupont1995phase,atanasov1996coherent, hache1997observation, potz1998coherent, rioux2011current, baranova1993multiphoton, baskin1988coherent, sipeol1991}. Coherent currents, resulting from the coherent photogalvanic effect (PGE), have been utilized to induce effective second-order nonlinear susceptibility ($\chi^{(2)}$) in glasses with a technique called all-optical poling (AOP). During AOP, the phase dependence of coherent currents allow the inscription of periodically sign alternating $\chi^{(2)}$ gratings by coupling strong fundamental harmonic pump to the waveguide. The latter leads to quasi-phase-matching (QPM) for momentum conservation amongst the involving photons and efficient energy conversion in nonlinear optical interactions. Such automatic QPM was intensely studied, predominantly for second-harmonic generation (SHG), both theoretically \cite{baskin1988coherent, sipeol1991, zel1989interference, baranova1990polar, dianov1991photoinduced} as well as experimentally  \cite{osterberg1986dye, margulis1995imaging, dianov1995photoinduced, Tom:88, Krol:91, margulis1989phase} in optical fibers more than two decades ago. However, the link between theory and experiment, as well as the influence of the material parameters remained mostly unexplored. 

In recent years, AOP was also demonstrated in integrated photonics, both in waveguides \cite{billat2017large, hickstein2019self, nitiss2019formation} and microring resonators \cite{ lu2021efficient, nitiss2021optically}, and photoinduced second-order nonlinearity regained significant attention. Integrated photonics offers improved modal confinement allowing high intensities under reduced powers compared to optical fibers and increased flexibility for dispersion engineering. Particularly, stoichiometric silicon nitride (Si$_3$N$_4$) with its large transparency window, low losses, high third-order susceptibility ($\chi^{(3)}$), high refractive index and mature nanofabrication process is very appealing. It is exploited for several linear \cite{blumenthal2018silicon} and nonlinear applications, such as four-wave mixing \cite{kruckel2015continuous,ayan2022polarization}, third-harmonic generation (THG) \cite{levy2011harmonic}, supercontinuum \cite{grassani2019mid} and Kerr comb \cite{gaeta2019photonic} generation using $\chi ^{(3)}$ nonlinearity. AOP now allows to add  $\chi^{(2)}$ processes, such as  SHG, difference-frequency generation \cite{sahin2021difference} and spontaneous parametric down-conversion \cite{dalidet2022near}, to the already impressive nonlinear toolbox of Si$_3$N$_4$. Recently, there have been several qualitative attempts to explain the time dynamics of AOP, which exhibits growth and saturation, and its dependence on waveguide dimensions \cite{hickstein2019self,nitiss2019formation,nitiss2020highly}. However, a quantitative assessment is still lacking with unknown material constants, while the seeding mechanism of the process remains elusive \cite{terhune1987second, dianov1991photoinduced, stolen1990second,stolen1987self,chmela1988second, Lu2021Kerr}. With two-photon microscopy (TPM) imaging, charge gratings have been observed \cite{hickstein2019self}, providing crucial information in terms of interacting modes. However, the length of the grating always remained shorter than half of the waveguide length \cite{hickstein2019self, nitiss2019formation} suggesting that the light conversion efficiency could be further increased. Overall, the origin and physical limitations of AOP remain unclear. A possible approach for quantitative analysis, and therefore a mean to optimize efficiencies, is to use a seeded approach where a coherent second-harmonic (SH) is externally coupled together with its pump. With such a seeded approach, the required time and powers for AOP were significantly reduced in fibers \cite{Krol:91}, while the efficiencies could be increased. As such, AOP is initiated in a more efficient and controlled fashion.

In this work, based on the model initially proposed by Dianov et al \cite{dianov1991photoinduced}, we develop a general phenomenological model that predicts the existence of non-degenerate sum-frequency generation (SFG) QPM gratings and set the basis for explaining the dynamics of AOP in waveguides. Our model is experimentally validated with the first observation of SFG QPM gratings initiated by the coherent interaction of a pump and its SH in a Si$_3$N$_4$ waveguide. In addition, we formulate the time dynamics of the SHG process enabled by coherent PGE. The dynamic model serves as a basis for the extraction of material parameters critical in the inscription of the $\chi^{(2)}$ gratings, which now provides a mean to benchmark the performances of different material platforms. In our case, we apply this approach to Si$_3$N$_4$ through a series of experiments, relying on a seeded AOP scheme. Finally, we show that such information is essential and can be used in establishing the net conversion efficiency limitation and energy requirements of the AOP process, thus setting an important building block in the optimization of such devices. The remainder of the paper is organized as follows: the first section describes the general phenomenological model for the coherent PGE and the experimental validation. The second section focuses on the development of the equations governing the dynamic formation of the space-charge grating and the subsequent extraction of the phase of the process, the photogalvanic coefficient and photoconductivities in Si$_3$N$_4$. The last section relates these extracted material parameters to the achievable SHG efficiencies of a given platform.

\section{Photoinduced generation of coherent current in centrosymmetric media}\label{sec:quantum}
\begin{figure}[h!]
    \centering
    \includegraphics[width=\linewidth]{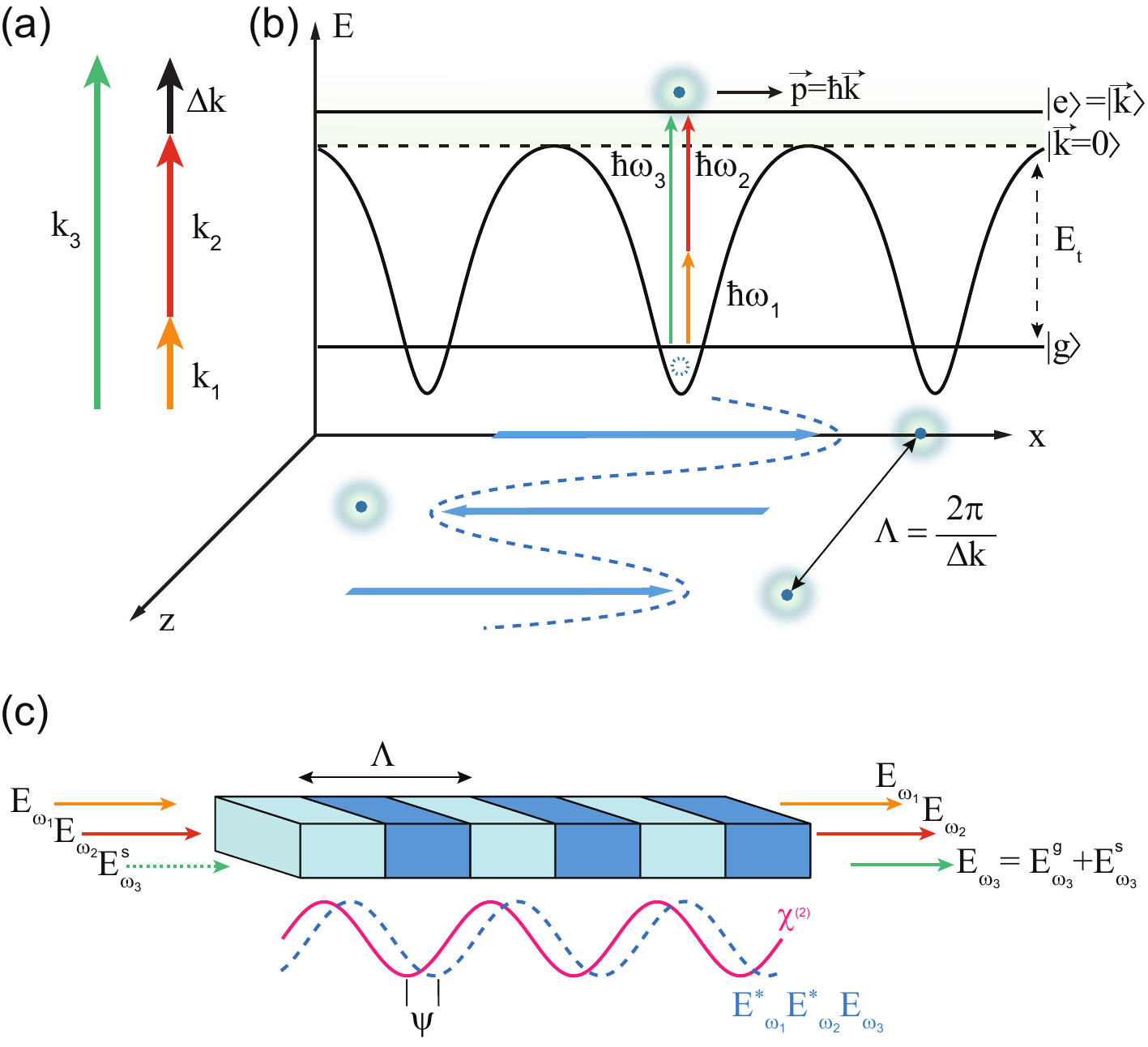}
    \caption{(a) The phase matching diagram for SFG QPM gratings. (b) Charge transport mechanism in AOP for SFG. The directional movement of charges ($\vec{p}=\hbar \vec{k}$) are due to the interference of one photon (angular frequency $\omega_3$, wavevector $k_3$) and two photons (frequency $\omega_1$ and $\omega_2$, wavevectors $k_1$ and $k_2$) ionization pathways from trap $\ket{g}$ with $E_t$ being the depth of the trap to excited state $\ket{e}$. The movement is periodic along the propagation direction and is dependent on the wavevectors mismatch $\Delta {k}={k}_3-{k}_1-{k}_2$ of incident waves resulting in a space-charge field  with period $\Lambda$.  (c) The schematic of optically inscribed QPM gratings with period $\Lambda$ is shown. There can be a phase shift $\psi$ between the inscribed $\chi^{(2)}$ grating and the product of the participating fields.}
    \label{fig:SFGpge}
\end{figure}

\subsection{Phenomenological model}
In this section, we analyze the movement of a trapped charge carrier that is exposed to coherently related fields yielding multiphoton absorption interference. We start by defining our model in which under an unperturbed Hamiltonian $\mathbb{H}_0$ the charge carrier is trapped in ground state $\ket{\Psi _g(t)}$ and can be excited to the conduction band $\ket{\Psi _e(t)}$ via absorption of coherently related photons with optical wavevectors of $k_1$, $k_2$ and $k_3$ and angular frequencies $\omega_1$, $\omega_2$ and $\omega_3$ satisfying $\omega_3 = \omega_1 + \omega_2$ as shown in Fig. \ref{fig:SFGpge}. Here $\ket{\Psi _n(t)}=\ket{n}e^{-i\omega_n t}$ ($n=\{g,e\}$). The charge carrier excitation to conduction band by optical fields enables currents inside the material.
We assume the interacting field as phase-locked, and its vector potential, $\vec{\mathcal{A}}$, stated as
\begin{equation}
\vec{\mathcal{A}} = \sum_{j=1}^3\vec{\mathcal{A}}_{\omega_j}e^{i(k_jz-\omega_j t)}+c.c.~,
\end{equation}
where $k_j$ ($j=\{1,2,3\}$) are wavevectors of the absorbed photons, and $c.c.$ stands for complex conjugate. In the model we will assume that the light fields are polarized along the x direction while propagating along z-axis.
It is important to note that, as shown in the energy diagram in Fig. \ref{fig:SFGpge}, the charge carrier can be excited with a single photon at $\omega_{3}$, while it requires two photons with lower energy, $\omega_{1}$ and $\omega_{2}$, respectively, for ionization of the same charge carrier.
Because the traps are deep ($E_{t}/kT\gg 1$), it is assumed all carriers are initially in the trap ($a_e^{(0)}=0$). When a carrier in the trap state is subject to light, the interaction potential in the long wavelength limit is $\mathbb{V}=-\frac{q}{m}\vec{\mathcal{A}}\cdot \vec{\mathbb{p}}$, where q, m and $\mathbb{p}$ are the charge, mass and momentum operator, respectively. We are interested in the one dimensional problem so we will drop the vector signs in the potential. 
The coefficients for single photon absorption ($a^{(1)}_e$) of the sum-frequency (SF) photon of frequency $\omega_3=\omega_1+\omega_2$ and two photon absorption ($a^{(2)}_e$) of frequencies $\omega_1$ and $\omega_2$ are acquired using first- and second-order perturbation theory, respectively. Then the probability amplitude of the excited state is $a_e \approx a^{(0)}_e +a^{(1)}_e + a^{(2)}_e$ and is proportional to \cite{landau2013quantum}:
\begin{equation}
a_e \sim p_{eg} \mathcal{A}_{\omega_3} + \gamma p_{ei}p_{ig}  \mathcal{A}_{\omega_1} \mathcal{A}_{\omega_2}~,
\end{equation}
where $\gamma$ is constant, the momentum holds the relation $p_{ab}=\braket{a|\mathbb{p}|b} $, $\ket{i}$ is the intermediate state. The momentum has an odd parity in position, and the charge density that is related to $|a_e|^2$ normally contains even powers of momentum (or $k$). Therefore, under absorption, the charge density change does not give rise to a directional current. However, it becomes uneven in momentum space when the power of $k$ is uneven in the cross terms of $|a_e|^2$ \cite{atanasov1996coherent, baranova1990polar}. This leads to an asymmetry in ionization due to interference of one- and two-photon dipole moments \cite{dupont1995phase}. Using that the electric field $E=-\frac{\partial \mathcal{A}}{\partial t}$, i.e., $E_{\omega_j}=i\omega_j \mathcal{A}_{\omega_j}$, the ionization rates parallel ($\dot{\rho}^+$) and anti-parallel ($\dot{\rho}^-$) to the vector potential become $\dot{\rho}^{\pm} \sim |E_{\omega_3}\pm i\zeta E_{\omega_1}E_{\omega_2}|^2$, where $E_{\omega_i}$ is the electric field amplitude of light of frequency $\omega_i$, and $\zeta$ a constant linked to the ionized charge's momentum and absorption cross-sections.

Such anisotropic excitation leads to a photocurrent $j_{ph} \sim \dot{\rho}^+ - \dot{\rho}^-$. Including the higher-order perturbations from deeper traps, we can write the photocurrent as $j_{ph}=(\beta_0 + \beta_{100} |E_{\omega_1}|^2+\beta_{010} |E_{\omega_2}|^2+ \beta_{001} |E_{\omega_3}|^2+...)E_{\omega_3}E_{\omega_1}^*E_{\omega_2}^*e^{i\Delta kz}e^{-i\psi}$, where $\psi$ is the shift between the inscribed grating and the product of the participating fields ($E_{\omega_3}E_{\omega_1}^*E_{\omega_2}^*$) at any given point in the waveguide (Fig. \ref{fig:SFGpge}(b)), and $\beta_{ijk}$ is the photogalvanic coefficient found from $(i+j+k+2)^{th}$ perturbation, $\Delta k=k_3-k_1-k_2$ is the unmatched wavevectors of absorbed photons. As described in Fig. \ref{fig:SFGpge}(b), the contributing light at $\omega_3$ consists of both the seed ($E_{\omega_3}^{\rm s}$) and the generated wave from the inscribed grating ($E_{\omega_3}^{\rm g}$). The excitation of charge carriers also leads to increase in conductivity $\sigma=\sigma_{001} I_{\omega_3} +\sigma_{110}I_{\omega_1}I_{\omega_2}+\sigma_{101}I_{\omega_1}I_{\omega_3}+\sigma_{011}I_{\omega_2}I_{\omega_3}+\sigma_{002}I_{\omega_3}^2+...$, where $I_{\omega_i}$ is the intensity of light of frequency $\omega_i$ and $\sigma_{ijk}$ is the photoconductivity coefficient of $i+j+k$ photon absorption where the subscript denotes the number of photons absorbed from frequencies $\omega_1$, $\omega_2$ and $\omega_3$, respectively. Considering only the second-order perturbation, we can simplify the expressions for the induced photocurrent and conductivity as Eqs. \eqref{eq:jph} and \eqref{eq:sigma}, respectively.
\begin{equation}
j_{ph}=\beta E_{\omega_3}E_{\omega_1}^*E_{\omega_2}^*e^{i\Delta kz}e^{-i\psi}~,
\label{eq:jph}
\end{equation}
\begin{equation}
\sigma=\sigma_{001} I_{\omega_3}+\sigma_{110}I_{\omega_1}I_{\omega_2}~.
\label{eq:sigma}
\end{equation}

As can be seen both the anisotropic current and conductivity depend nonlinearly on the optical fields and intensities. In a final steady state, the charge separation leads to the inscription of a static electric field ${E}_{DC}$:
\begin{equation}
E_{DC}=-\frac{{j}_{\textit{ph}}}{\sigma}~,
\label{eq:EDC}
\end{equation}
where the sign of the current is spatially modulated in the x-z plane (Fig. \ref{fig:SFGpge}) along the light propagation direction due to unmatched wavevector $\Delta k$ of absorbed photons. The momentum mismatch leads to the spatial modulation of DC field with a period $\Lambda=2\pi/\Delta k$.

\subsection{Experimental validation}

\begin{figure}[h!]
    \centering
    \includegraphics[width=\linewidth]{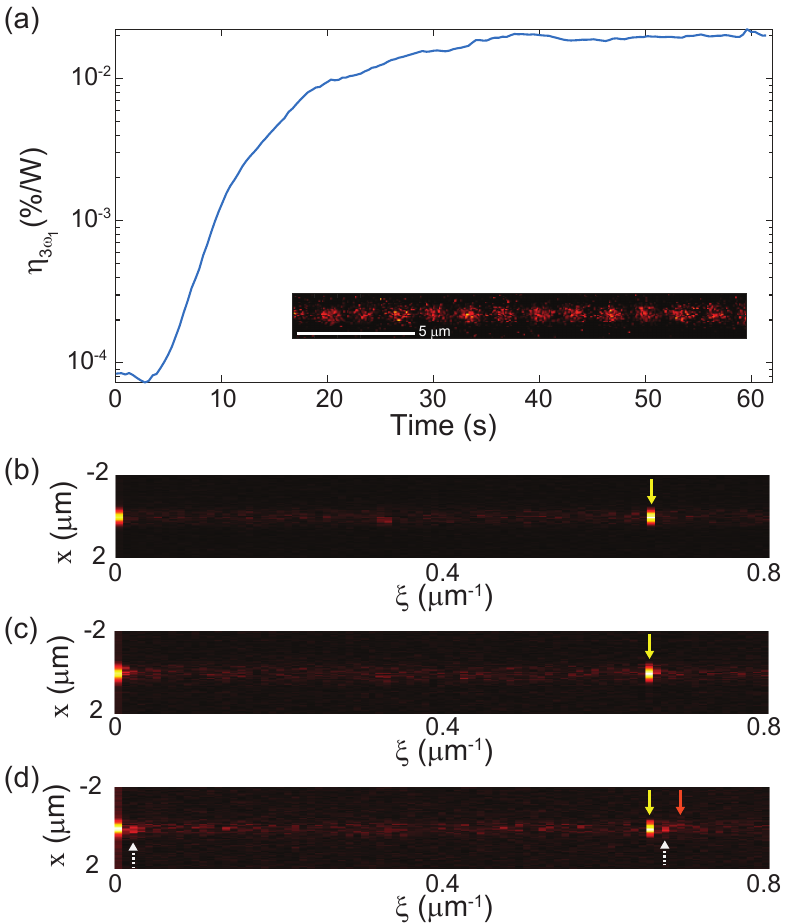}
    \caption{(a) Growth of TH (SF) conversion-efficiency during AOP with the introduction of pump and its externally generated SH in time. The average outcoupled pump power is $43.6$ mW while the SH is $0.3$ mW.  Inset: TPM image of the $1^{st}$ meander of the waveguide. Spatially-resolved Fourier analysis of TPM images of $\chi^{(2)}$ gratings in (b) $1^{st}$, (c) $6^{th}$ and (d) $9^{th}$ meander of 9 cm long waveguide folded in 9 meanders with cross-section of $0.57 {\rm \,\mu m} \times 0.81 {\rm \, \mu m}$. The yellow arrow identifies the primary grating for SHG, the orange arrow - the secondary grating for SFG, and the dashed white arrows - the interference between the two.}
    \label{fig:sfg}
\end{figure}

While our general model predicts the existence of SFG QPM gratings, all coherent PGE demonstrations so far were for the degenerate SHG case, where $\omega_{1}$=$\omega_{2}$=$\omega_{3}/2$. To validate our model we used the experimental setup described in Appendix \ref{sec:setup}: a $1.55~{\rm \mu m}$ pump ($\omega_{1}$) and its SH ($\omega_{2}$=2$\omega_{1}$), shaped in ns pulses, are simultaneously coupled to a 9 cm long Si$_3$N$_4$ waveguide folded in meanders and with $0.57 {\rm \,\mu m} \times 0.81 {\rm \,\mu m}$ cross-section. It was previously observed \cite{nitiss2020highly} that coupling only the pump light results in the spontaneous growth of its SH power at the output of waveguide. However, when both pump and its SH are coupled, we can also clearly observe the spontaneous growth of third-harmonic (TH). The THG CE defined as $\eta_{3\omega_1}=P_{3\omega_1}/(P_{2\omega_1}P_{\omega_1})$ is shown in Fig. \ref{fig:sfg}. As the grating period is related to the wavevector mismatch of the participating optical waves, those can be revealed from processing the grating images capture by TPM imaging \cite{nitiss2021optically}. Such an image can be seen in the inset of Fig \ref{fig:sfg}(a). In order to precisely retrieve the grating components, we perform a spatially resolved Fourier analysis on grating images recorded at different positions along the waveguide length. The results for gratings on the $1^{st}$, $6^{th}$ and $9^{th}$ meanders are shown in Figs. \ref{fig:sfg}(b-d), respectively. 

We see in the first meanders a single non-zero spatial frequency (yellow arrow) which corresponds to the grating phase-matched to the interference of pump-SH absorption, having wavevector mismatch of $\Delta k_1 = k_{2\omega_1}-2k_{\omega_1}$. In the last meander, additional non-zero spatial frequencies appear owing to the inscription of a secondary grating due to the interference of pump-SH-TH absorption, assumed to have wavevector mismatch of $\Delta k_2= k_{3\omega_1}-k_{\omega_1}-k_{2\omega_1}$ (orange arrow). Considering that the final $\chi^{(2)}$ grating is composed of two parts, the measured $(\chi^{(2)})^2$ response should possess four nonzero spatial frequencies proportional to  $2|\Delta k_1|$, $2|\Delta k_2|$, and interference terms $|\Delta k_1 + \Delta k_2|$ and $|\Delta k_1 -\Delta k_2|$. In our case, the secondary grating being weak, its spatial frequency does not clearly appear but can still be easily retrieved based on the interference terms (dashed white arrows). 
The experimentally obtained spatial frequency related to $2|\Delta k_1|$ and inferred $2|\Delta k_2|$ are in agreement within $3 \%$ errors with those obtained from finite element method simulations, considering pump-SH and pump-SH-TH photogalvanic processes involving dominantly fundamental mode interactions. These findings unambiguously confirm the possibility of having non-degenerate case of AOP as well as the simultaneous inscription of multiple $\chi^{(2)}$ gratings.

\section{TIME DYNAMICS OF ALL-OPTICAL POLING AND ITS DEPENDENCE ON WAVEGUIDE MODES} \label{sec:dynamics}
\subsection{Theoretical bases}
In this section, we present the governing equation for the formation of space-charge gratings under coherently related fields, focusing on the SHG enabled by coherent PGE. In this case, we denote the pump frequency as $\omega_1$, and thus the seeded SH is at frequency of $2\omega_1$. We can rewrite Eq. \eqref{eq:jph} as $j_{ph}=\beta E_{2\omega_1}(E_{\omega_1}^*)^2e^{i\Delta k_1z}e^{-i\psi}$ and  Eq. \eqref{eq:sigma} as $\sigma=\sum_{a,b}\sigma_{ab}I_{\omega_1}^{a}I_{2\omega_1}^{b}$ for $a\geq 2$ and $b\geq0$, where $\beta$ is a function of intensities of pump and SH due to the contributions of higher-order perturbations or involvement of other states, i.e., $\beta=\sum_{a,b} \beta_{ab} I_{\omega_1}^{a}I_{2\omega_1}^{b}$. We start by solving the continuity equation and Maxwell’s equations together under slowly-varying envelope and undepleted pump approximation \cite{dianov1991photoinduced} (see Appendix \ref{sec:shg}). Introducing the walk-off between the pump and SH with a term $g(z)=e^{-(z/L_{\rm w})^2}$, where $L_{\rm w}$ the temporal walk-off length is a function of group-velocity mismatch of involved modes (See Appendix \ref{sec:walkoff}), as $|E_{\omega_1}|\to |E_{\omega_1}|g(z)$ \cite{Tom:88}, we obtain:
\begin{equation}
    \frac{\partial^{2} \bar{E}}{\partial t \partial z}-\frac{\omega_1 \chi^{(3)}}{2n_{2 \omega_1} c \epsilon} \beta |E_{\omega_1}|^4 e^{i(\frac{\pi}{2}-\psi)} g^4(z) \bar{E}=0~,
    \label{governing}
\end{equation}
where $\bar{E}=E_{2\omega_1} e^{t/\tau}e^{\alpha z/2}$ with the relaxation time $\tau=\epsilon/\sigma$, $\epsilon$ being the dielectric constant of the material, and $\alpha$ is the propagation loss coefficient of SH. 

The optical modes participating in the grating inscription should also be examined. We use the separability of the modes in different dimensions, i.e., $E^{(w)}_{q\omega_1}(\vec{r},t)=$ $U^{(w)}_{q\omega_1}(x, y) A^{(w)}_{q\omega_1}(z,t)$ ($q=1,2$), where $U^{(w)}_{q\omega_1}$ describes the normalized transverse field and $A^{(w)}_{q\omega_1}$ is the field amplitude along the z-axis of the $w^{th}$ mode. In addition, we multiply both sides of the Eq. \eqref{governing} with $\iint dxdy\, (U^{(w)}_{2 \omega_1}(x,y))^*$ and define normalization $\iint dxdy\, |U^{(w)}_{q \omega_1}(x,y)|^{2}=1$ to obtain
\begin{equation}
    \frac{\partial^{2} A^{(l)}}{\partial t \partial z}=M_p \Gamma_{pl} A^{(l)}-\sum_{a,b} \frac{e^{\alpha b z}|A_{\omega_1}^{(p)}|^{2a}|A^{(l)}|^{2b}}{\tau^{{\rm eff}}_{ab}(S_{\omega_1}^{(p)})^a (S_{2\omega_1}^{(l)})^b} \frac{\partial A^{(l)}}{\partial z},
    \label{eq:nlgov}
\end{equation}
where $M_p=\frac{i\omega_1\chi^{(3)}}{2n_{2\omega_1}c\epsilon}\beta|A^{(p)}_{\omega_1}|^4e^{-i\psi}$, $S_{\omega_1}^{(p)}$ and $S_{2\omega_1}^{(l)}$ are the effective areas of pump and SH modes as defined in the Appendix \ref{sec:QPM}. $\tau^{\rm eff}_{ab}=\frac{\epsilon}{\sigma_{ab}\kappa_{ab}}\frac{2^{a+b}}{(\epsilon_0 c)^{a+b} n_{2\omega_1}^bn_{\omega_1}^a}$, $\kappa_{ab}$ is defined as $S_{\omega_1}^{(p)^a} S_{2\omega_1}^{(l)^b}\iint dxdy\, |U_{\omega_1}(x,y)|^{2a} |U_{2 \omega_1}(x,y)|^{2b+2}$, $A^{(l)}=A^{(l)}_{2\omega_1}e^{\alpha z/2}$ and we define the overlap integral of modes for $C_{4 v}$ point group as
\begin{equation}
    \Gamma_{pl}=\iint dxdy\, |U^{(p)}_{\omega_1}(x,y)|^{4}|U^{(l)}_{2 \omega_1}(x,y)|^{2}~,
    \label{eq:overlap}
\end{equation} 
where the symmetries of third-order susceptibility and nonlinear conductivity is hidden here. It describes the effective nonlinear overlap of the modes (unit of $m^{-4}$) that write a grating and generate SH. Here, $A^{(l)}$ is independent of area of the waveguide as the power transferred in the mode $l$ is $\frac{n_{2\omega_1} c \epsilon_0}{2}|A^{(l)}_{2\omega_1}|^2$. The information on the waveguide dimensions are hidden in the overlap integral. The overlap integral is in the same order of magnitude for different modes and can be simulated numerically (see Appendix C). Hence, the inscription of QPM gratings of different modes is allowed as observed in other works \cite{yakar2021seeded,nitiss2021optically}. The same applies to the more generalized SFG process.

As evident, the governing Eq. \eqref{governing} encloses important material constants that should be addressed in order to quantify the capabilities of a given platform: $\beta$ will determine the magnitude of the coherent photocurrent, which will be counterbalanced by the increase in conductivity $\sigma$ owing to the excitation of charge carriers. Another interesting term is the phase shift $\psi$ between the inscribed grating and, in the case of SHG, the SH at any given point in the waveguide. In most analyses of the coherent PGE in optical waveguides, it is assumed to be $\pi/2$, which is justified by the fact that coherent current is caused by the $\pi/2$ phase difference in the single and two-photon ionizations \cite{baranova1993multiphoton, anderson1991model}. However, the phase of the photocurrent can be different if the influence of the atomic potential is considered \cite{anderson1992interference, baranova1993multiphoton, baranova1991physical}. 
In the following subsections we experimentally extract $\beta$, $\sigma$ and $\psi$ that will provide means for quantitative assessment of the capabilities of Si$_3$N$_4$ platform for SHG.

\subsection{Experimental extraction of $\psi$, $\beta$ and $\sigma$ in Si$_3$N$_4$ waveguides} \label{sec:pm}

\begin{figure}[htb]
    \centering
    \includegraphics[width=\linewidth]{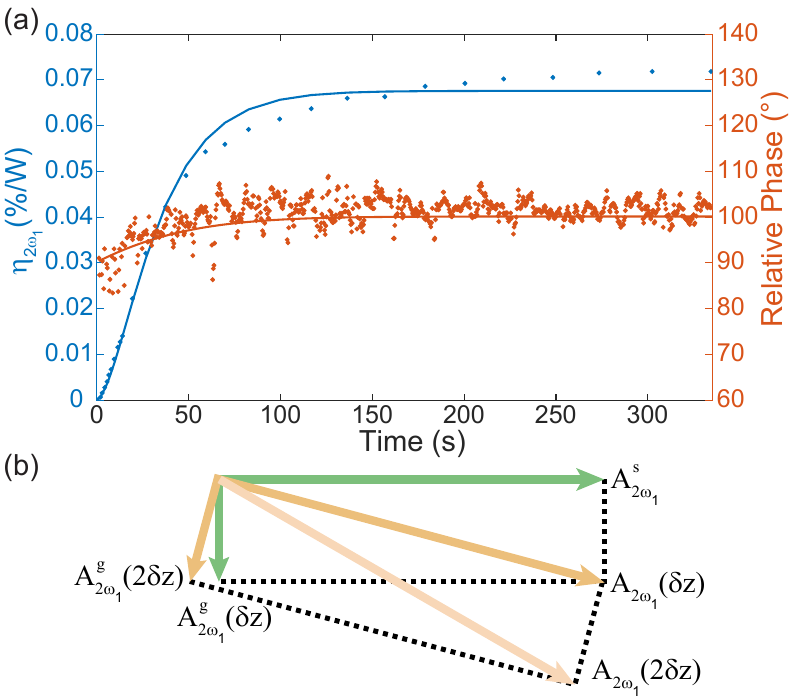}
    \caption{(a) CE and relative phase of the generated SH with respect to the seed phase during AOP. Dots are experimental data and lines are fits. The CE was fitted using Eq. \eqref{Aanalytical} for $\psi=0$. (b) Evolution of the total SH amplitude and phase along the waveguide for $\psi=0$.}
    \label{fig:phasemes}
\end{figure}
Parametric optical and photogalvanic processes can have different phase shifts \cite{stolen1990second,anderson1992interference, baranova1993multiphoton, baranova1991physical}. In ref. \cite{margulis1989phase}, the steady state phase of the generated SH with respect to the seed SH was measured to be $\pi/2$ for silica fibers. From that, it was concluded the phase of inscribed $\chi^{(2)}$ grating is the same as the product of the participating fields $(E_{\omega_1}^*)^2E_{2\omega_1}$, i.e., $\psi=0$. In order to extract the dynamic phase of the grating for the Si$_3$N$_4$ waveguide, we seed the AOP process and measure the interference between the seed and the generated SH. We use a 43 mm long waveguide with $1.5\,{\rm \mu m} \times 0.8\,{\rm \mu m}$ cross-section, and inject pump and SH with peak powers of $8.57$ W and $400$ mW, respectively. Without the loss of generality, we consider the field amplitude of SH seed at the input of the waveguide writes as $A_{2\omega_1}^{\rm s}$ with an initial phase of 0, the generated SH field amplitude is of the form $A^{\rm g}_{2\omega_1} e^{i\varphi}$. If we consider the process at the beginning of the waveguide, the input SH seed ($A_{2\omega_1}^{\rm s}$) leads to generated SH $A_{2\omega_1}^{\rm g}(\delta z)$ after a distance $\delta z$. Clearly if the SH generated is strong enough to contribute to the subsequent grating inscription, i.e.  $A_{2\omega_1}^{\rm s}$+$[A_{2\omega_1}^{\rm g}(\delta z)]$ becomes the new seed, this grating will be shifted by a phase related to the strength of $A_{2\omega_1}^{\rm g}(\delta z)$. The same then occurs throughout the length of the waveguide and it is expected that the grating will thus have a varying phase. This is illustrated in Fig. \ref{fig:phasemes}(b), for the specific case of $\psi=0$. As Eq. \eqref{governing} is symmetric in position and time, analogous evolution of fields happen in time as well. In the case of high seed power, the grating inscription is dominated by the seed SH such that the contribution of the internally generated SH can be considered negligible. Under this limit, we can minimize the phase evolution and extract $\psi$ from the generated SH phase by $\varphi = \pi/2 - \psi$. The CE is defined as $\eta_{2\omega_1} =P_{2\omega_1}/P_{\omega_1}^2= \frac{2n_{2\omega _1}}{\epsilon_0 c n_{\omega _1}^2} \frac{|A^{\rm g}_{2\omega_1}|^2}{|A_{\omega_1}|^4}$ and experimentally retrieved phase $\varphi$ during the poling process are shown in Fig. \ref{fig:phasemes}. Out of 90 poling events of 4 different waveguides, we extract $\psi=-1.9^{\circ}\pm14.8^{\circ}$ (or similarly $181.9^{\circ}\pm14.8^{\circ}$).  We attribute the fluctuations to the weak generated SH in the beginning of the process reducing the resolution of the interference measurements. Similar to \cite{margulis1989phase}, we can conclude that the total generated SH and the inscribed grating are in phase, i.e. $\psi=0$ in our case. 

We extract the material parameters $\beta$ and $\sigma$ through a series of experiments. 
To link results to theory, we make certain assumptions and design the experiments accordingly. We first assume that the dark conductivity is very small compared to the photoconductivity \cite{nitiss2019formation, madou2018fundamentals, tuncer2017nonlinear}. Therefore, we can ignore the grating erasal due to the dark conductivity. In order to linearize the equation, we assume the pump and SH powers are constant except for the linear losses. This assumption can be justified once again in the case  of high seed power since it will dominate the total SH in the poled waveguide. Using such approximations we solve Eq. \eqref{governing}  (See Appendix \ref{sec:shg}), to get the second-order susceptibility and SH field. For the generalized case of a constant pump and seed in the $p^{\rm th}$ and $l^{\rm th}$ mode respectively, the solution is expressed as
\begin{widetext}
\begin{subequations}
\begin{align}
\chi^{(2)}_{pl}(\vec{r}, t)=\frac{\chi^{(3)}}{\epsilon} \beta\left(E_{\omega_1}^{{(p)}*}(x,y)\right)^{2} E_{2 \omega_1}^{{\rm s},(l)}(x,y) e^{i(\Delta k_1 z-\psi)} g^2(z) e^{-\alpha z / 2} \int_{0}^{t} d t^{\prime} J_0\left(2 \sqrt{-M_p \Gamma_{pl} G(z) t^{\prime}}\right) e^{-t^{\prime} / \tau} ~,\label{chi2analytical}\\
A^{{\rm g},(l)}_{2 \omega_1}(z, t)=-A_{2 \omega_1}^{{\rm s},(l)} e^{-\alpha z / 2} \int_{0}^{2 \sqrt{-M_p \Gamma_{pl} G(z) t}} d r e^{r^{2} / 4 M_p \Gamma_{pl} G(z) \tau} J_{1}(r) ~.\label{Aanalytical}
\end{align}
\end{subequations}
\end{widetext}
It can be seen that the $\chi^{(2)}$ grating adjusts its shape to the product of involved modes, i.e. $\left(E_{\omega_1}^*\right)^{2} E_{2 \omega_1}^{\rm s}$ and is periodic with a period $\Lambda=2\pi/\Delta k_1$. Under this configuration, $\beta$ and $\sigma$ can be extracted by fitting the time dynamics of the SHG CE with Eq. \eqref{Aanalytical}. We also assume that the process occurs dominantly on fundamental modes for both pump and SH, as previously shown. Hence, for the remainder of the paper we will not denote the mode number in overlap integrals and amplitudes. The optical setup and procedures for the measurements are described in Appendix \ref{sec:setup}. 

\begin{figure}[h!]
    \centering
    \includegraphics[width=\linewidth]{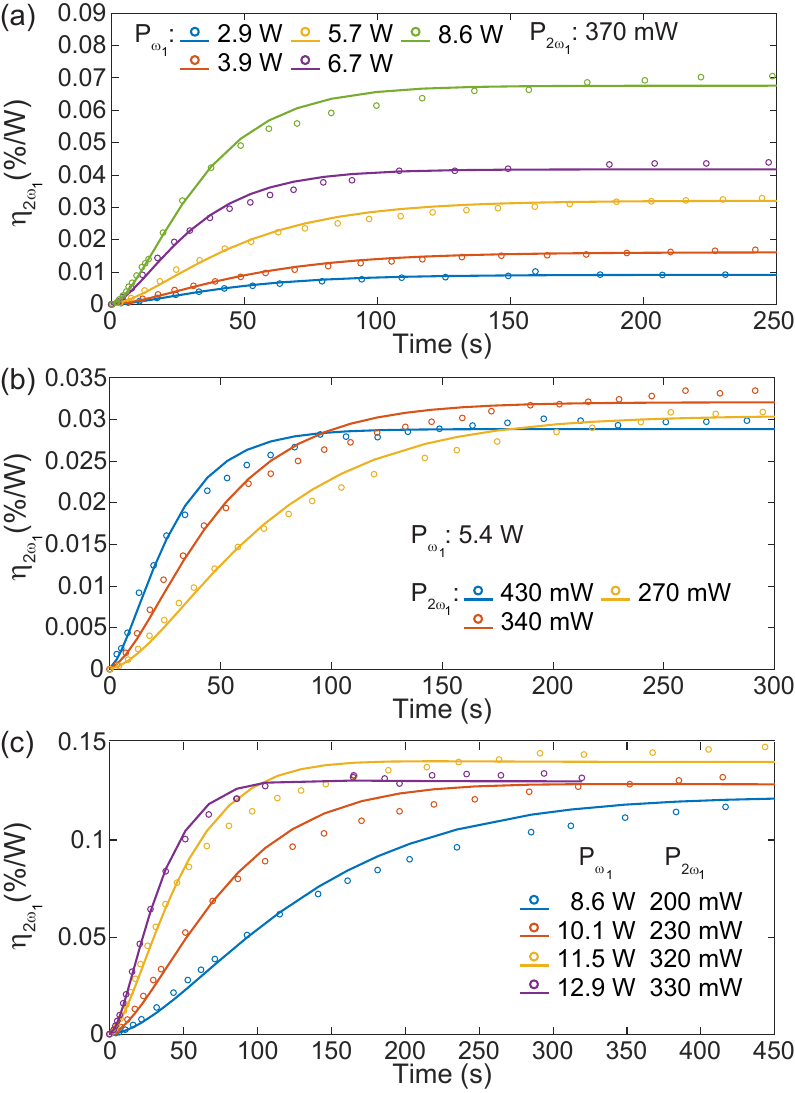}
    \caption{Experimental SHG CE during AOP (dotted line) and fit (solid line) for (a) constant peak SH seed power of 66.2 mW and varying peak pump power; (b) constant peak pump power of 3.4 W and varying peak SH seed power; (c) varying peak power for both pump and SH seed.}
    \label{fig:fitpol}
\end{figure}

We carried out AOP on four 43 mm long waveguides with different cross-sections ($1.1$, $1.3$, $1.4$, $1.5$ $\rm \mu m$ $\times$ $0.8$ $\rm \mu m$), and for various coupled pump and coupled SH seed powers. The extracted CE data is then fitted using Eq. \eqref{Aanalytical} and examples of the experimental data and fit for the $1.5$ ${\rm \mu m}$ $\times$ $0.8$ ${\rm \mu m}$ cross-section waveguide are shown in Fig. \ref{fig:fitpol}. We can point out that the power requirement to initiate AOP is significantly reduced with seeding compared to the spontaneous process (i.e. only pump is injected in the waveguide) \cite{nitiss2019formation}. We can see in Fig. \ref{fig:fitpol}(a) that the initial growth rate and efficiency of the process increases with pump power. When the pump power is fixed and we vary the seed SH power a similar behavior is observed in terms of speed however the reached efficiency does not significantly vary but seems to start decreasing with the highest seed power. This is confirmed in Fig. \ref{fig:fitpol}(b). As the photocurrent depends on the $(A_{\omega_1}^2)^* A_{2\omega_1}$, the initial growth rate increases with the increase of the product of the coupled power ($P^2_{\omega_1}P_{2\omega_1}$) but the efficiency is limited as the photoconductivity increases. This is demonstrated in Fig. \ref{fig:fitpol}(c). One can see that similar efficiencies can be reached for lower pump and SH.

The photogalvanic coefficient and photoconductivity can be intensity dependent due to the higher-order perturbations or contribution of deeper states. Hence, using the data from 90 experimental poling events of the four waveguides, we fit $\beta$ and $\sigma$ with a polynomial as a function of coupled pump and SH intensities. The fits, presented in Table \ref{tab:matconst}, were obtained using least absolute residuals method with $R^2=0.92$ and $R^2=0.93$, for $\beta$ and $\sigma$, respectively. The extracted conductivity is orders of magnitude higher than the dark conductivity values found in literature \cite{madou2018fundamentals}. 

\begin{table}[htb]
    \centering
    \caption{Extracted photoconductivity and photogalvanic coefficients}
        \label{tab:matconst}
        \begin{ruledtabular}
        \begin{tabular}{c|c|c|c}
          $\sigma_{10}\,{\rm (S \mu m/W)}$ & $\sigma_{01}\,{\rm (S \mu m/W)}$ &$\sigma_{02}\,{\rm (S \mu m^3/W^2)}$& $\sigma_{21}\,{\rm (S \mu m^5/W^3)}$  \\[3pt] \hline
          $\approx0$ & $\approx0$  & $1.19\times10^{-15}$ & $1.36\times10^{-18}$ \\[2pt]\hline \hline
          $\beta_{0}\, {\rm (A\mu m /V^3)}$ & $\beta_{10}\, {\rm (\mu m^3 /V^4)}$ & $\beta_{01}\, {\rm (\mu m^3 /V^4)}$ & $\beta_{20}\, {\rm (\mu m^5 /V^5A)}$ \\[3pt] \hline
         $\approx0$  & $\approx0$  &  $1.25\times10^{-19}$  & $9.38\times10^{-23}$  \\[2pt]
    \end{tabular}
    
    \end{ruledtabular}
\end{table}

\begin{figure}[h!]
    \centering
    \includegraphics[width=\linewidth]{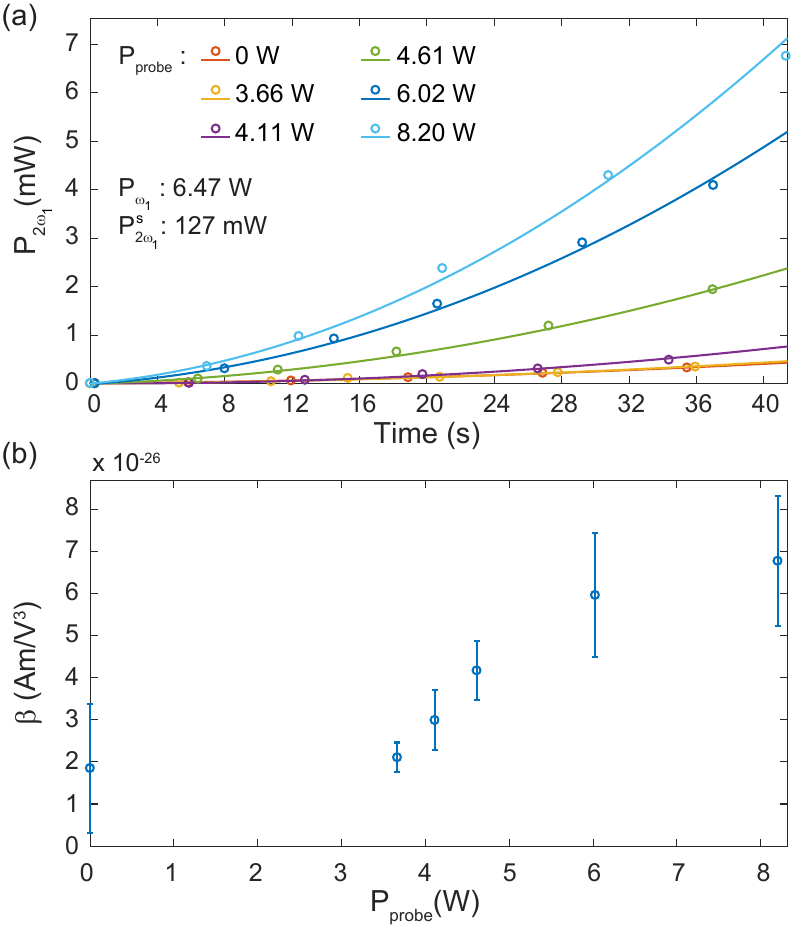}
    \caption{(a) Generated SH power as a function of time for the probe light of varying peak powers. During the different poling events, the forward propagating pump and SH powers are kept constant. Dots are experimental data, lines are fits. The waveguide has a cross section of $1.79\times 0.8\,{\rm \mu m}^2$. (b) Extracted $\beta$ from fit as a function of probe power.}
    \label{fig:incoh}
\end{figure}

The fact that our experimental data yields $\sigma_{01}$ close to zero  raises the question of possible involvement from intermediate states or higher-order of the coherent PGE. In order to gain insight on this, we modified the experimental setup to check if the higher-order contributions to current and conductivity come from higher-order coherent PGE \cite{sipeol1991} or from incoherent photo-excitation from deeper traps working as a carrier source \cite{dianov1993growth}. To that end, we split the pump beam and send part of the pump beam backwards (probe) in addition to the forward propagating pump and its SH (see Fig. \ref{fig:setup}) inside waveguide having cross-section $1.79\,{\rm \mu m}\times0.8\,{\rm \mu m}$. We ensured that forward and backward pump and probe pulses are coincidental. The forward propagating pump and SH beams work as coherent sources and the backward propagating beam as the incoherent source. In Fig. \ref{fig:incoh}, we sweep the incoherent pump power keeping the coherent sources constant, similar to the work in ref. \cite{dianov1993growth}, and observe the increase of the initial speed of the process with increase of incoherent counter-propagating pump power. We extract $\beta$ by making a parabolic fit to the power using the Taylor approximation of Eq. \eqref{Aanalytical}. The increase in $\beta$ as a function of backward propagating probe, and hence increase in the photocurrent from the forward propagating pump and SH, is an evidence of deeper traps working as a charge carrier source. The backward probe promotes carriers to intermediate state, thus increasing the number of carriers which can then be involved in the third-order coherent PGE as shown in Fig. \ref{fig:SFGpge}. While higher-order coherent PGEs may also explain such behavior, as pointed out in \cite{dianov1993growth}, such effects are much less probable. Our observations are therefore in agreement with the three-photon model and involvement of intermediate states.

\section{Conversion Efficiency and Performance Limitations}\label{sec:experiment}
\begin{figure*}[htb]
    \centering
    \includegraphics[width=1\linewidth]{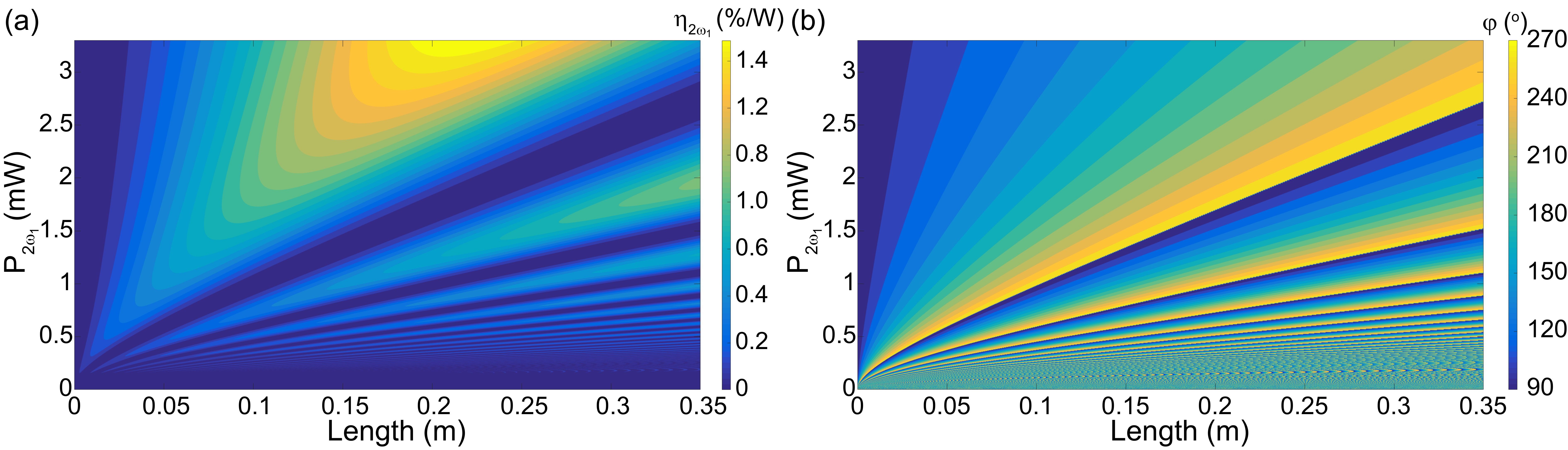}
    \caption{Estimated CE ($\rm \%/W$) for a waveguide of cross-section $1.4\,{\rm \mu m} \times 0.8 \, {\rm \mu m}$ (a) and phase of the generated SH ($\varphi$) (b) along the length and average seed power is shown for $\psi=0$. Simulation is done for the material constants in Tab. \ref{tab:matconst} and average pump power of $0.1$ W. For a certain length, the maximum conversion efficiency can be optimized with the introduction of external seed power.}
    \label{fig:Contour}
\end{figure*}
The measured material parameters now provide means for making a quantitative study of achievable SHG CE in the Si$_3$N$_4$ platform. It is important to emphasize, that due to the phase parameter $\psi$ which in our case is 0, the output generated SH and the input seed SH may have a phase difference as illustrated in Fig. \ref{fig:phasemes}(b). The phase variation depends on waveguide length, initial SH seed power as well as SH power generated inside the waveguide, and it is expected that the grating will thus have a varying phase. 

By analytically solving the governing Eq. \eqref{governing} in a steady state, we obtain the amplitude and phase of SH generated inside the waveguide after AOP (see Eqs. \eqref{eq:Assincoh} and \eqref{eq:phissincoh} in the Appendix \ref{sec:shg}). The simulated relative amplitude and phase of the generated SH as a function of seed amplitude and waveguide length are plotted in Figs. \ref{fig:Contour}(a,b), respectively. For high seed powers, the grating inscription is dominated by the seed field and the phase change is relatively slow along the length. For low seed powers where generated light power is comparable with the seed light power, the grating inscription is influenced by both the seed and the generated field. There are oscillations in the phase and generated field amplitude along the length. As the seed power reduces, the phase and efficiency fluctuations become more pronounced. This is due to the one part of the grating that is interfering destructively with another part of the grating,  as $\psi$ is not equal to $\pm \pi/2$. Another limiting factor comes from the increase of the photoconductivity with increased seed power. For high seed powers, conductivity takes over the photocurrent and the effective $\chi^{(2)}$ reduces, as seen from Eq. \eqref{eq:EDC}, reducing the efficiency. However, as the phase change is small the achievable CE rises with the length for high seed powers where grating inscription is dominated by the seed light. From the simulations, it is observed that the achievable CE rises over $1.4$ ${\rm \%/W}$ for $20$ cm long waveguide with the high seed powers. Despite the trade-off between seed power and DC field for high seed powers, further efficiencies can be reached for longer waveguides with higher seed powers.

\section{Conclusions}

In this work, we extend the phenomenological model proposed by Dianov et al. \cite{dianov1991photoinduced} to understand the underlying physics and dynamics of the AOP process in waveguides, and to develop physical bases to compare different platforms. We observe the dynamic phase for Si$_3$N$_4$ and extract $\psi$ to be approximately $0$ or $\pi$. For the first time, we find a solution for the governing equation at high seed approximation and, therefore, we retrieve the photogalvanic coefficient $\beta$ and photoconductivity $\sigma$ for Si$_3$N$_4$ and make predictions of the expected efficiencies. We can realize that with seeding, the energy requirements can be reduced by an order of magnitude and the speed increased drastically compared to the spontaneous AOP. In addition, we predict and provide the first experimental demonstration of  the inscription of $\chi^{(2)}$ gratings for the general case of SF generation.

\begin{acknowledgments}
This work was supported by ERC grant PISSARRO (ERC-2017-CoG 771647). The samples used for the experiment were fabricated by LIGENTEC SA.
\end{acknowledgments}

\appendix

\section*{Appendix}

The Appendix includes: the details on the experimental setup, derivations of equations used in the main text, numerical simulations of overlap integral and temporal walk-off length for all-optical poling, material parameters that are used in estimating the values in Tab. \ref{tab:matconst} and the decay curves of the gratings under illumination. 

\section{Experimental setup} \label{sec:setup}

\begin{figure}[htb]
    \centering
    \includegraphics[width=\linewidth]{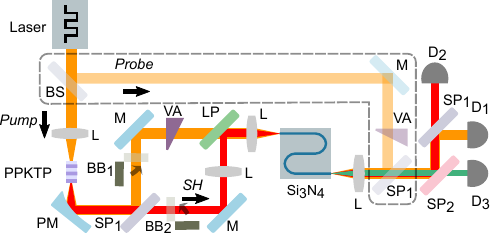}
    \caption{Seeding setup for AOP of SiN waveguides. Pulsed light is sent to PPKTP after which SH and pump are collimated in the parabolic mirror. Then, the chromatic aberration between the pump and SH of is compensated using an interferometer like setup and an optical lens in SH beam path. Light is coupled to the TE mode of the waveguide. The chip is thermally stabilized using a temperature controlling stage. The semi-transparent part of the schematic (enclosed in a dashed box) is added for the experiments to determine the order of coherent PGE. BS: Beam splitter, L: Lens, PM: Parabolic mirror, SP: Short-pass dichroic mirror (1: cutoff at $1100$ nm, 2: cutoff at $650$ nm), BB: Beam block, VA: variable optical attenuator, LP: Long-pass dichroic mirror, D: Detector. }
    \label{fig:setup}
\end{figure}

The schematic of setup used for this work is shown in Fig. \ref{fig:setup}. Light from a tunable continuous wave laser at the wavelength of $1.55~{\rm \mu m}$ is shaped into $1$ ns square pulses with duty cycle of $1/200$, and then amplified. For the demonstration of SFG and the extraction of the nonlinear parameters, the pump is guided through a type-0 periodically poled KTP crystal (PPKTP) to coherently generate SH. Before being coupled to waveguide with a lens, pump and SH take different paths as to enable their independent control. As such their powers can be adjusted while we use beam blockers BB$_1$ and BB$_2$ to selectively pass only SH and pump, respectively. The inverse tapers at the facets of the waveguides enable the efficient coupling of pump and SH into the fundamental modes of the waveguides. We compensate the chromatic aberration of pump and SH and increase light coupling efficiency via an interferometer-like setup with an optical lens in the SH beam path. The maximum available on-chip pump and SH peak powers are $12.9$ W and $744$ mW, respectively, yet those can be varied separately using variable optical attenuator (VA). In such a configuration, strong pump and SH seed can be introduced which is crucial for the demonstration of AOP in an SF generation process. The output of the chip is collected, and the pump, SH and TH components are separated via dichroic mirrors and sent to detectors. Our setup also enables coupling of the probe light from the output port as to study the incoherent contribution of pump light to the AOP process. In this case, the amplified laser output, before doubling, is split into two equal parts labeled as pump and probe. It is important to note that the optical path length of pump and probe is equalized to guarantee the temporal overlap of optical pulses throughout the length of the waveguide during AOP process.

The measurement of the CE during the AOP is performed as follows. The seeded SH is blocked for a quick instant at certain times using BB$_{2}$ and the generated SH power is recorded. This is performed repeatedly until saturation is observed (see for example Fig. \ref{fig:fitpol}). After each experiment, the space-charge grating is erased by blocking pump with BB$_{1}$ and coupling just the SH. Similarly, the time decay of the generated SH is measured by closing BB$_2$ and opening BB$_1$ for an instant. 

For the phase measurements of the generated SH during AOP, we use the fact that the seed, generated and total SH are bound by the law of cosines:

\begin{equation}
    \varphi=\pm {\rm acos}\left(\frac{|A_{2 \omega _1}|^2-|A_{2 \omega _1}^{\rm s}|^2-|A_{2 \omega _1}^{\rm g}|^2}{2{|A_{2 \omega _1}^{\rm s}||A_{2 \omega _1}^{\rm g}|}}\right) + 2m\pi~,
    \label{eq:varphi}
\end{equation}
where m is an integer, $A_{2 \omega _1}$, $A_{2 \omega _1}^{\rm s}$, $A_{2 \omega _1}^{\rm g}$ are the total, seed and generated SH amplitudes and $\varphi$ is the phase difference between the seed and generated SH.

The generated SH is fitted using Eq. \eqref{Aanalytical}. For the fit, AOP is assumed to occur just during the pulse duration. The peak to average power ratios of the pump and seed SH are experimentally characterized for this purpose. Due to the limited extinction ratio of the intensity modulator, the ratio is measured to be around $130$ for pump. While at SH, the peak power converges to $200$-times of the average power owing to the peak enhancement in the nonlinear crystal. 

\section{Solving the Second-Harmonic Generation Equation} \label{sec:shg}
\subsection{Derivation of Dynamic Equation}

Here, we derive the governing SH generation equation for a second-order nonlinearity caused by coherent PGE as in ref. \cite{dianov1991photoinduced}. As shown in Section II of the main text, the coherent PGE leads to generation of currents. Writing the continuity equation for the total current in the waveguide ($\vec{j}_{tot} = \vec{j}_{ph} + \sigma \vec{E}_{DC}$), we have
\begin{equation}
\frac{\partial \rho}{\partial t} = - \vec{\nabla}.\vec{j}_{tot}=- \vec{\nabla}.(\vec{j}_{ph} + \sigma \vec{E}_{DC})~.
\end{equation}
Using Gauss' Law, and canceling the divergences on both sides, one gets 
\begin{equation}
\epsilon  \frac{\partial \vec{E}_{DC}}{\partial t} = - \vec{j}_{ph} - \sigma \vec{E}_{DC}~,
\end{equation}
 where $\epsilon=\epsilon_0 \epsilon_{r,DC}$ with $\epsilon_0$ and $\epsilon_{r,DC}$ being the permittivity of the free space and relative permittivity of the medium, respectively. Photocurrent and the current due to the DC field are in opposite directions meaning that the inscribed field in the material will contribute to the relaxation of separated charges. Using $\chi^{(2)} = \chi^{(3)}E_{DC} $, and solving Maxwell’s equations under slowly-varying envelope and undepleted pump approximations \cite{fejer1992quasi}, we get:
\begin{subequations}
\label{eq:wholesetap}
\begin{align}
    \frac{\partial \chi^{(2)}}{\partial t} &= \frac{\chi^{(3)}}{\epsilon}j_{ph} - \frac{\chi^{(2)} }{\tau}~, \label{eq:continuity} \\
\frac{\partial E_{2\omega_1}}{\partial z} &= \frac{i \omega_1}{2n_{2\omega_1}c}\chi^{(2)}E^2_{\omega_1}e^{-i\Delta k z} - \frac{\alpha}{2} E_{2\omega_1} ~,\label{eq:shg}\\
j_{ph} &= \beta (E^*_{\omega_1})^2 E_{2\omega_1} e^{i (\Delta k_1 z - \psi)}+c.c. ~,\label{eq:photocurrent}\\
\tau & =\epsilon/\sigma ~,\label{eq:tau}
\end{align}
\end{subequations}
where $\Delta k_1=k_{2\omega_1}-2k_{\omega_1}$, $n_{2\omega_1}$ and $\alpha$ are the effective refractive index and linear loss at frequency of SH respectively and $\psi$ is the constant phase shift determined by the phenomenological model discussed in Section \ref{sec:quantum} and atomic potentials.

We can simplify Eqs. \eqref{eq:wholesetap} for a constant $\tau$, first defining $\bar{\chi}=\chi^{(2)}e^{t/\tau}e^{\alpha z/2}$ and $\bar{E}=E_{2\omega_1} e^{t/\tau}e^{\alpha z/2}$. So, Eqs. \eqref{eq:continuity} and \eqref{eq:shg} become
\begin{subequations}
\begin{align}
\frac{\partial \bar{\chi}}{\partial t} &=\frac{\chi^{(3)}}{\epsilon} \beta\left(E_{\omega_1}^{*}\right)^{2} \bar{E} e^{i(\Delta k_1 z-\psi)}~,\label{dchi2dt}\\
\bar{\chi} &=-i \frac{2n_{2 \omega_1} c}{\omega_1} \frac{1}{E_{\omega_1}^{2}} \frac{\partial \bar{E}}{\partial z} e^{i \Delta k_1 z}~. \label{chi2shg}
\end{align}
\end{subequations}
Taking the partial derivative of Eq. \eqref{chi2shg} and plugging into Eq. \eqref{dchi2dt}, we obtain
\begin{equation}
    \frac{\partial^{2} \bar{E}}{\partial t \partial z}-\frac{\omega_1 \chi^{(3)}}{2n_{2 \omega_1} c \epsilon} \beta |E_{\omega_1}|^4 e^{i(\frac{\pi}{2}-\psi)} \bar{E}=0 ~,\label{b5}
\end{equation}
where $M=\frac{i\omega_1\chi^{(3)}}{2n_{2\omega_1}c\epsilon}\beta|E_{\omega_1}|^4e^{-i\psi}$ is defined for clarity. We can treat $M$ as a constant for a constant pump. We also introduce a walk-off function between the pump and SH ($g(z)$) as $|E_{\omega_1}|\to |E_{\omega_1}|g(z)$ \cite{Tom:88}. By taking the Laplace transform, Eq. \eqref{b5} rewrites, 
\begin{equation}
    \partial_z(e^{-MG(z)/s}\mathcal{L}[\bar{E)}])=\frac{1}{s}e^{-MG(z)/s}\partial_z\bar{E}(z,0)~,
    \label{governingapp}
\end{equation}
where $\mathcal{L}[\bar{E)}](z,s)$ is the Laplace transform of $\bar{E}(z,t)$ and $G(z)=\int_0^{z}dz^\prime g^4(z^\prime)$. Assuming there is no $\chi^{(2)}$ in the beginning of the AOP process, we have $\partial_z\bar{E}(z,0)=0$. Therefore, one obtains 
\begin{equation}
    e^{-MG(z)/s}\mathcal{L}[\bar{E)}]=C(s)~.
\end{equation}
For a constant seed $E_{2\omega_1}(0,t)=E_{2\omega_1}^{\rm s} u(t)$, where u is the unit-step function, we have $\bar{E}(0,s)=E_{2\omega_1}^{\rm s}/(s-\tau^{-1})$, hence $\bar{E}(z,s)$ is obtained. The $\mathcal{L}[\bar{\chi}]$ becomes, 
\begin{equation}
    \mathcal{L}[\bar{\chi}] = \frac{\chi^{(3)}\beta}{\epsilon}\left(E_{\omega_1}^*\right)^2 E_{2\omega_1}^{\rm s} g^2(z) \frac{e^{MG(z)/s}}{s(s-\tau^{-1})} e^{i (\Delta k_1 z-\psi)}~.
    \label{eq:chift}
\end{equation}
We can go back to $\mathcal{L}[{\chi^{(2)}}](z,s)=\mathcal{L}[\bar{\chi}](z,s+\tau^{-1})e^{-\alpha z/2}$. In addition, using $\mathcal{L}[J_0(2\sqrt{at})]=\frac{e^{-a/s}}{s}$ \cite{abramowitz1948handbook}, we obtain Eq. \eqref{chi2analytical}. When the seed is closed for a moment, using Eq. \eqref{eq:shg}, the generated SH amplitude gives Eq. \eqref{Aanalytical}. In addition, to have the most general solution, the effect of initial and intrinsic $\chi^{(2)}$ and solution in sec. \ref{sec:shg} can be superimposed. However, throughout the text we ignore the contribution of intrinsic $\chi^{(2)}$ as it is not phase matched and the phase difference between electronic and optical processes cancels the contribution of intrinsic $\chi^{(2)}$ to the grating inscription.

\subsection{Spectral Properties of the Poled Waveguides} \label{sec:QPM}

Under the high seed approximation and ignoring the loss, we find the steady state value of Eq. \eqref{eq:chift} as
\begin{equation}
    \chi^{(2)} (\vec{r}) = \frac{\chi^{(3)}\beta}{\sigma}\left(E_{\omega_1}^*\right)^2 E_{2\omega_1}^{\rm s} g^2(z) e^{M\Gamma G(z)\tau} e^{i (\Delta k_1 z-\psi)}~.
    \label{eq:chi2ss}
\end{equation}

\begin{figure}[h!]
     \centering
     \includegraphics[width=1\linewidth]{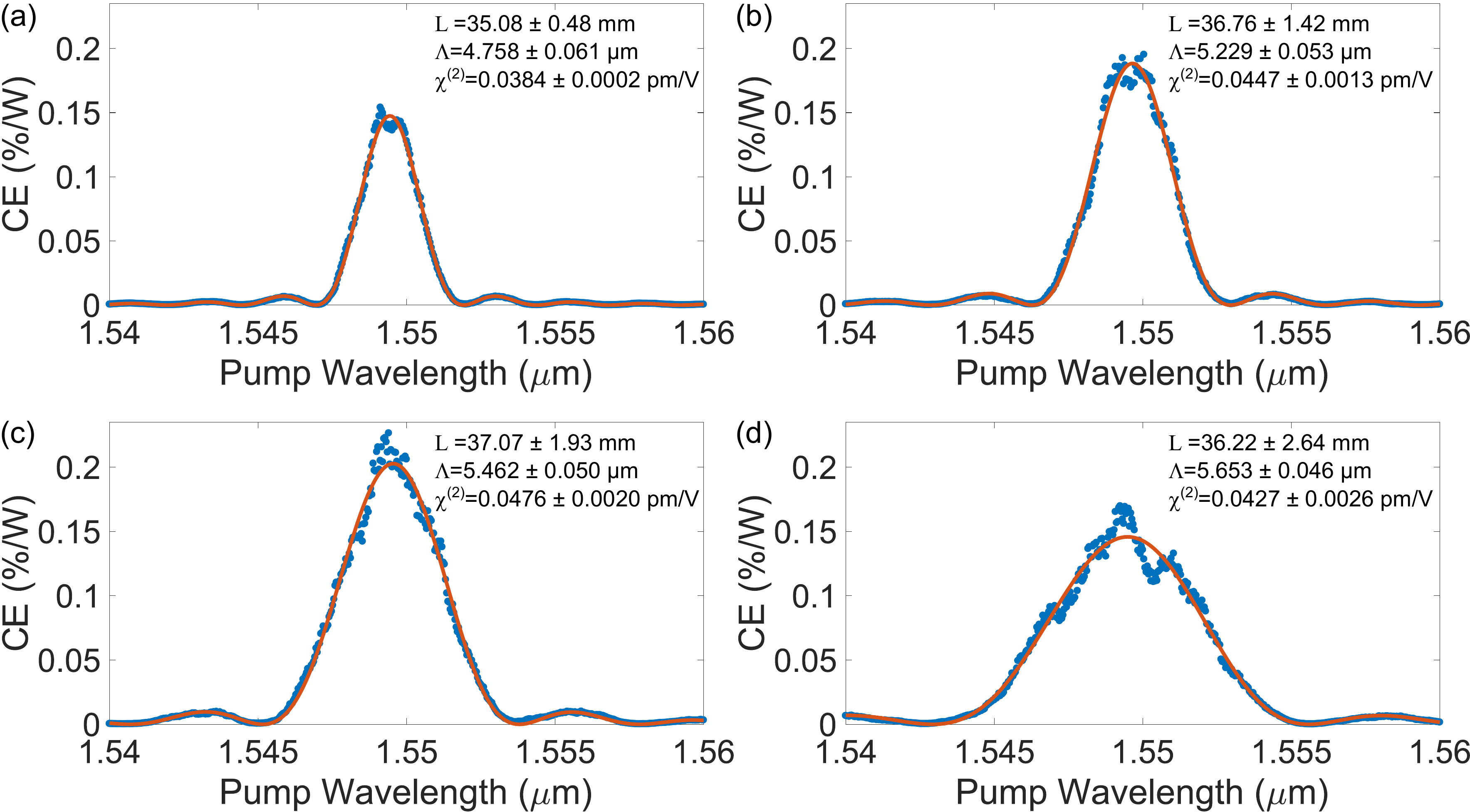}
     \caption{(a)-(d) Measured CE spectra (blue dots) after AOP at $1.55\, {\rm \mu m}$ with their fits (red line) of waveguides with cross-sections $1.1$, $1.3$, $1.4$ and $1.5$ $\rm \mu m$ $\times$ $0.8$ $\rm \mu m$, respectively. The fitting parameters are effective second-order nonlinearity $\chi^{(2)}_{\rm eff}$, grating period ($\Lambda$) and length ($L$) are presented in the inset.}
     \label{fig:Spectralfit}
 \end{figure}

Here, one can see that in the seeded case the grating length is expected to be equal to the waveguide length in the case seed and pump are constant as $\psi=0$, $M$ is purely imaginary so it does not change the amplitude of the grating along the waveguide. Similar to refs. \cite{lin2007nonlinear,lu2021efficient,nitiss2021optically}, we define $\Gamma = \bar{\eta}/\bar{S}^2$, where $\bar{\eta}$ is the normalized overlap integral and effective area is defined as $\bar{S}= (S_{\omega_1}^2 S_{2\omega_1})^{1/3}$, where 
\begin{equation}
    S_{\omega_q}=\frac{1}{\left(\iint dx dy |U_{\omega_q}|^6\right)^{1/2}}~.
\end{equation}
Hence, effective second-order nonlinearity can be defined as $\chi^{(2)}_{eff}=\frac{\chi^{(3)}\beta}{\sigma}\left(A_{\omega_1}^*\right)^2 A_0^{\rm s}\frac{\bar{\eta}}{\bar{S}^{3/2}}$. After AOP process, the grating period is $\Lambda=2\pi/\Delta k_1$. Integrating the SHG equation (Eq. \eqref{eq:shg}) along the cross-section, one obtains 
\begin{equation}
    \frac{\partial A_{2\omega_1}}{\partial z} = \frac{i \omega_1}{2n_{2\omega_1}c}\frac{\chi^{(2)}_{\rm eff}}{\sqrt{\bar{S}}} A^2_{\omega_1}e^{-i(\Delta k_1-\frac{2\pi}{\Lambda}) z} e^{-i\psi}~.
\end{equation}
Plugging Eq. \eqref{eq:chi2ss} into above equation, integrating it, and using $\psi=0$ or $\pi$, one obtains
\begin{equation}
    \eta_{2\omega_1}= \frac{(\omega_1 \chi^{(2)}_{\rm eff} L)^2}{2c^3\epsilon_0\bar{n}^2 \bar{S}} {{\rm sinc}^2\left( (\Delta k_1-{\tfrac{2\pi}{\Lambda}}-|M|\Gamma \tau){\tfrac{L}{2}}\right)}~,
\end{equation}
where $\bar{n}=(n_{2\omega}n_\omega^2)^{1/3}$, L is the waveguide length. For the waveguides and powers used $2\pi/\Lambda\gg|M|\Gamma \tau$. Hence, one can ignore the latter term. Thus, we can find the grating length from the spectral fits just using the simulated waveguide dispersion. The experimental CE spectrum and its fit are shown in Fig. \ref{fig:Spectralfit}. We find the grating length close to the waveguide length from the fits and the grating images. Hence, we justify the assumption of grating length being equal to the waveguide length.

\subsection{Solving the steady state equation}
We can analytically solve the nonlinear governing equation (Eq. \eqref{eq:nlgov}) for the steady state for ignoring the losses and the walk-off between the pump and SH, i.e. $G(z)\approx z$. For the steady state the time derivative cancels. So, we have 

\begin{equation}
    M \Gamma =\frac{1}{\tau_{0}^\prime A} \frac{\partial A}{\partial z}+\kappa \frac{A^*}{\tau_{2}} \frac{\partial A}{\partial z}~.
\end{equation}

We integrate both sides with respect to z and we define $A=|A|e^{i\phi}$. We get

\begin{eqnarray}
 M \Gamma z =&& \frac{1}{\tau_{0}^\prime} \left(\ln\left(\frac{|A|}{|A_0^{\rm s}|}\right)+i\phi\right)\nonumber \\
 && +\kappa \left(\frac{|A|^2-|A_0^{\rm s}|^2}{2\tau_2}+i\frac{\phi |A|^2}{\tau_2}\right)~.
\end{eqnarray}

We can separate the real and imaginary parts and solve it. Thus, we define $M=|M|e^{i\theta}$, where $\theta=\frac{\pi}{2}-\psi$. We have
\begin{equation}
    |M| \Gamma z \cos\theta = \frac{1}{\tau_{0}^\prime} \ln\left(\frac{|A|}{|A_0^{\rm s}|}\right)+\kappa \frac{|A|^2-|A_0^{\rm s}|^2}{2\tau_2}~,
\end{equation}
\begin{equation}
    |M| \Gamma z \sin\theta =\phi\left( \frac{1}{\tau_{0}^\prime} +\kappa \frac{|A|^2}{\tau_{2}} \right)~.
\end{equation}

Solving both equations we have

\begin{equation}
    |A|=\sqrt{\frac{\tau_2}{\kappa\tau_0^\prime}W\left(\frac{\tau_0^\prime\kappa|A_0^{\rm s}|^2}{\tau_2}e^{\tau_0^\prime(\kappa\frac{|A_0^{\rm s}|^2}{\tau_2}+2|M|\Gamma z\cos \theta)}       \right)}~,
    \label{Ass}
\end{equation}
\begin{equation}
    \phi=\frac{|M|\Gamma z \sin \theta}{\frac{1}{\tau_{0}^\prime} +\kappa \frac{|A|^2}{\tau_{2}}}~,
    \label{phiss}
\end{equation}
where W is Lambert W function. These solutions are sum of seed and generated SH. To get the generated SH, one can simply subtract the seed amplitude ($A_0^{\rm s}$) from $A$. 

\subsection{Solving the steady state equation for general photocurrent and photoconductivity}
In this subsection  we solve the Eq. \eqref{eq:nlgov} for the steady state including the incoherent contributions for the nonzero terms in Table \ref{tab:matconst}. Similar to the previous derivation, we start with setting time derivatives of Eq. \eqref{eq:nlgov} to 0 and ignoring the losses ($\alpha=0$) and we have 
\begin{eqnarray}
M_p \Gamma_{pl} A^{(l)}&&=\sum_{a,b} \frac{|A_{\omega_1}^{(p)}|^{2a}|A^{(l)}|^{2b}}{{\tau^{{\rm eff}}_{ab}}S_{\omega_1}^{(p)^a} S_{2\omega_1}^{(l)^b}} \frac{\partial A^{(l)}}{\partial z}\nonumber\\
&&=\frac{1}{\tau}\frac{\partial A^{(l)}}{\partial z}~.
    \label{eq:nlgovap} 
\end{eqnarray}

We can separate the real and imaginary parts and solve it. Thus, we define $M=|M|e^{i\theta}$, where $\theta=\frac{\pi}{2}-\psi$. We also define $A=|A|e^{i\phi}$. We have

\begin{equation}
    \Gamma z \cos\theta = \int_{|A_0^{\rm s}|}^{|A|} \frac{dR^\prime}{|M|\tau R^\prime}~,
    \label{eq:Assincoh}
\end{equation}
\begin{equation}
    \phi = |M| \tau \Gamma z \sin\theta~.
    \label{eq:phissincoh}
\end{equation}

Here for $\psi=0$, and for the non-zero values of the tab. \ref{tab:matconst}, we have $|A|=|A_0^{\rm s}|$. As we represent $\phi$ in $[0,2\pi)$ region, there will be branch cuts, thus, there seems phase jumps in Fig. \ref{fig:Contour}. One simply subtracts the seed amplitude ($|A_0^{\rm s}|$) from $A$ to obtain $|A^{\rm g}_{2\omega_1}|e^{i\varphi}$.

\section{Temporal Walk-off Length for Photo-inscription of Charge Gratings} \label{sec:walkoff}

In this section, we present numerical calculations for the temporal walk-off length proposed in refs. \cite{Tom:88, weiner1998high}. The temporal walk-off length ($L_{\rm w}$) is defined as 
\begin{equation}
    L_{\rm w} = \left(\frac{c}{n_g(2\omega_1)-n_g(\omega_1)}\right)t_p~,
\end{equation}
where $t_p$ is the pulse length of pump, $n_g$ is the group index. The simulated walk-off length for all waveguides under test and for the fundamental TE modes at the pump and SH are above 20 m.  

The grating inscription cannot happen for distances above the walk-off length. All waveguides used are well below this limit justifying $G(z)\approx z$ approximation. 

\section{Material Parameters Used To Estimate The Coherent PGE Parameters}

In table \ref{tab:matparam}, we present the material constants that we used for Si$_3$N$_4$ using Eq. \eqref{Aanalytical}. The $\chi^{(3)}$ is obtained using $\chi^{(3)}=(4/3)n_{2\omega_1}\epsilon_0cn_2$ \cite{delCoso:04} and $n_2=2.5\times10^{-19}\, {\rm m^2W^{-1}}$ \cite{gaeta2019photonic}. 
\begin{table}[h!]
    \centering
        \caption{The material parameters used to extract the coherent PGE coefficients, i.e. $\beta$ and $\sigma$.}\label{tab:matparam} 
        \begin{ruledtabular}
    \begin{tabular}{c|c|c|c|c}
         $\lambda\,(\rm{nm})$ & $n_{2\omega_1}$ & $\epsilon_{r,DC}$ \cite{Sze2006Physics} &$\alpha\,(\rm{cm^{-1}})$ \cite{nitiss2020highly} & $\chi^{(3)}\,(\rm{m^2V^{-2}})$ \\\hline
          $1550$ & $2.0$ & $7.5$ & $0.168$ & $3.396\times10^{-21}$\\
    \end{tabular}
\end{ruledtabular}
\end{table}

\section{Overlap Integrals and Allowed Interactions} \label{sec:Overlap}

The Eqs. \eqref{eq:wholesetap} yield an overlap integral of modes presented in Eq. \eqref{eq:overlap} when the field interactions of pump modes and SH modes are considered. The overlap integral includes symmetries of both susceptibility and nonlinear conductivity. It is always non-zero intrinsically. This explains the AOP observed in higher-order modes of SH light in refs. \cite{lu2021efficient,nitiss2021optically,yakar2021seeded}. In this section, we will present the numerical values of overlap integrals using finite element method simulations implemented in \textit{Comsol Multiphysics} for different mode interactions and waveguide geometries.

\begin{table}[h!]
    \centering
        \caption{Overlap integrals ($\Gamma$) of first two modes of pump interaction with different TE SH modes for different waveguide widths for height of $0.8~{\rm \mu m}$.}\label{tab:Gamma} 
    \begin{ruledtabular}
    \begin{tabular}{c|c|c|c|c|c}
         $\Gamma_{1k}$ $({\rm \mu m^{-4}})$ &\textbf{SH$\bm{_1}$}&\textbf{SH$\bm{_2}$}&\textbf{SH$\bm{_3}$}&\textbf{SH$\bm{_4}$}&\textbf{SH$\bm{_5}$}\\\toprule
          $1.1$ ${\rm \mu m}$ & $2.15$ & $1.29$ & $1.25$ & $1.12$ & $0.90$\\
          $1.3$ ${\rm \mu m}$ & $1.78$ & $1.04$ & $1.04$ & $1.07$ & $0.61$\\
          $1.4$ ${\rm \mu m}$ & $1.62$ & $0.93$ & $0.94$ & $0.96$ & $0.55$\\
          $1.5$ ${\rm \mu m}$ & $1.47$ & $0.83$ & $0.85$ & $0.86$ & $0.49$\\
          $1.79$ ${\rm \mu m}$& $1.12$ & $0.63$ & $0.68$ & $0.65$ & $0.69$\\\toprule
          $\Gamma_{2k}$ $({\rm \mu m^{-4}})$ &\textbf{SH$\bm{_1}$}&\textbf{SH$\bm{_2}$}&\textbf{SH$\bm{_3}$}&\textbf{SH$\bm{_4}$}&\textbf{SH$\bm{_5}$}\\\toprule
          $1.1$ ${\rm \mu m}$ & $0.41$ & $0.46$ & $0.24$ & $0.35$ & $0.30$\\
          $1.3$ ${\rm \mu m}$ & $0.50$ & $0.68$ & $0.30$ & $0.52$ & $0.40$\\ 
          $1.4$ ${\rm \mu m}$ & $0.51$ & $0.72$ & $0.30$ & $0.54$ & $0.43$\\ 
          $1.5$ ${\rm \mu m}$ & $0.51$ & $0.74$ & $0.30$ & $0.54$ & $0.43$ \\
          $1.79$ ${\rm \mu m}$& $0.47$ & $0.73$ & $0.53$ & $0.27$ & $0.30$ \\
    \end{tabular}
    \end{ruledtabular}
        
\end{table}

Overlap integrals for different waveguide geometries are presented in Tab. \ref{tab:Gamma}. One infers the power requirements for grating inscription is higher as the dimensions get larger because the overlap integral is reducing for larger dimensions. However, overlap integrals of different modes are in the same order of magnitude, hence, interactions of different modes are allowed in the AOP process.  

\section{Erasal of Gratings Under Illumination} \label{sec:erasal}

There has been interest on the decay dynamics of the QPM gratings in fibers as it helps our understanding of the trap locations, time constants and physical mechanism of the process. The charge gratings decay faster at higher temperatures \cite{nitiss2019formation} or when exposed to high-energy photons \cite{dianov1993evidence, nitiss2019formation,sahin2021difference}. From Eq. \eqref{eq:continuity} and Table \ref{tab:matconst}, it is evident that the absence of pump causes a faster decay in second-order susceptibility as the photoconductivity increases with the introduction of SH. In this section, we show the erasure data of $\chi^{(2)}$ gratings with SH light and present the possible mechanisms behind photo-erasure. 

\begin{figure}[htb]
    \centering
    \includegraphics[width=\linewidth]{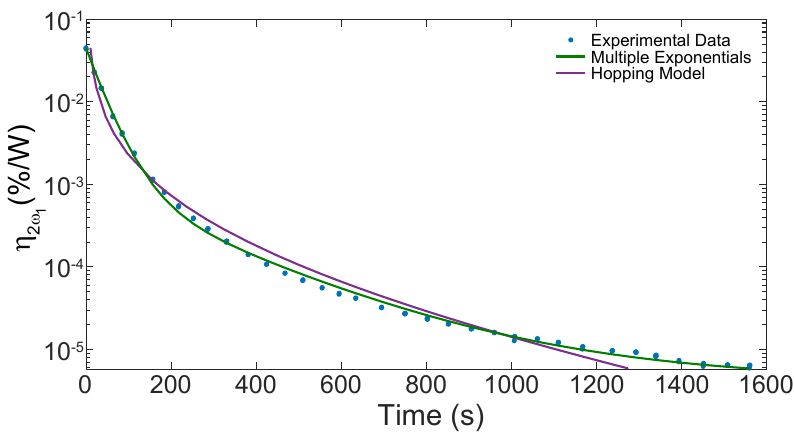}
    \caption{Decay curve of conversion efficiency along time under the illumination of pulsed SH light in the absence of pump. Fit is done using two different models in form of Eqs. \ref{eq:frankel} and \ref{eq:multipdef}.}
    \label{fig:diffmodels}
\end{figure}

The experiment was realized using the setup presented in Fig. \ref{fig:setup}. First, the pulsed SH light was coupled to the waveguide in the absence of pump and at certain times the conversion efficiency was measured in the absence of SH using just the pump beam. A decay curve in semi-logarithmic scale is shown in Fig. \ref{fig:diffmodels}. The conversion efficiency decay cannot be fitted with a single exponential function. There are several models to explain this. Charge hopping model \cite{nitiss2019formation} proposes the decay of second-order nonlinearity is of the form of 
\begin{equation}
    \frac{\partial\chi^{(2)}}{\partial t}=-K\chi^{(2)}\exp\left[L\sqrt{\chi^{(2)}}\right]~,
    \label{eq:frankel}
\end{equation}
where K and L are the fitting parameters, it has a solution of $t=N+K^{-1}E_1(L\sqrt{\chi^{(2)}})$ where $\rm{E}_1$ is the exponential integral \cite{abramowitz1948handbook}, $N$ depends on the initial conditions. 

Another model, also discussed in \cite{dianov1993evidence}, introduces slow and fast decay rates. In this case, the decay equation takes the form:
\begin{equation}
    \chi^{(2)}=C_1 e^{-t/\tau_1} + C_2 e^{-t/\tau_2} +D~,
    \label{eq:multipdef}
\end{equation}
where $C_1$ and $C_2$ coefficients of slow and fast decay and $D$ is the noise level and $\tau_1$ and $\tau_2$ are the time constants of the decay rates. In Fig. \ref{fig:diffmodels}, we present the conversion efficiency decay for the $1.5\, {\rm \mu m} \times 0.8\, {\rm\mu m}$ waveguide when SH light is coupled.  While thermal decay could be well fitted with the charge hopping model \cite{nitiss2019formation}, such model does not provide a good quality fit for the decay following optical illumination (Fig. \ref{fig:diffmodels}). The multiple exponential decay using Eq. \ref{eq:multipdef}, gives a good fit for the decay curve and the associated conductivity extracted from the fast decay rate is $2.4\times10^{-10}$ S/m for peak SH power of $326$ mW. We see the fast decay rate agrees with the fitted conductivity values in Table \ref{tab:matconst}.

\bibliography{apssamp}

\end{document}